\documentclass[conference]{IEEEtran}

\usepackage{graphicx}

\usepackage[style=ieee,natbib=true]{biblatex}
\usepackage{floatrow}
\usepackage[font=footnotesize,justification=centering]{caption}

\author{\IEEEauthorblockN{Nadine Boudargham\IEEEauthorrefmark{1},
Jacques Bou Abdo\IEEEauthorrefmark{2},
Jacques Demerjian\IEEEauthorrefmark{3}, Christophe Guyeux\IEEEauthorrefmark{4} and
Abdallah Makhoul\IEEEauthorrefmark{4}}
\IEEEauthorblockA{\IEEEauthorrefmark{1}Faculty of Engineering \\ Notre Dame University, Deir El Kamar, Lebanon\\ Email: nboudargham@ndu.edu.lb}
\IEEEauthorblockA{\IEEEauthorrefmark{2}Faculty of Natural and Applied Sciences \\ Notre Dame University, Deir El Kamar, Lebanon\\ Email: jbouabdo@ndu.edu.lb}
\IEEEauthorblockA{\IEEEauthorrefmark{3}LARIFA-EDST, Faculty of Sciences \\ Lebanese University, Fanar, Lebanon\\ Email: jacques.demerjian@ul.edu.lb}
\IEEEauthorblockA{\IEEEauthorrefmark{4}Femto-ST Institute, UMR CNRS 6174 \\ Universit\'e de Bourgogne Franche-Comt\'e, Besan\c{c}on, France \\Email: \{christophe.guyeux, abdallah.makhoul\}@univ-fcomte.fr}}

\title{Investigating Low Level Protocols for Wireless Body Sensor Networks}

\begin{document}

\maketitle

\begin{abstract}
The rapid development of medical sensors has increased the interest in Wireless Body Area Network (WBAN) applications where physiological data from the human body and its environment is gathered, monitored, and analyzed to take the proper measures. In WBANs, it is essential to design MAC protocols that ensure adequate Quality of Service (QoS) such as low delay and high scalability. This paper investigates Medium Access Control (MAC) protocols used in WBAN, and compares their performance in a high traffic environment. Such scenario can be induced in case of emergency for example, where physiological data collected from all sensors on human body should be sent simultaneously to take appropriate action. This study can also be extended to cover collaborative WBAN systems where information from different bodies is sent simultaneously leading to high traffic. OPNET simulations are performed to compare the delay and scalability performance of the different MAC protocols under the same experimental conditions and to draw conclusions about the best protocol to be used in a high traffic environment.
\end{abstract}


\section{Introduction}





\begin{figure}
\centering
\includegraphics[width=\columnwidth,height=\textheight,keepaspectratio]{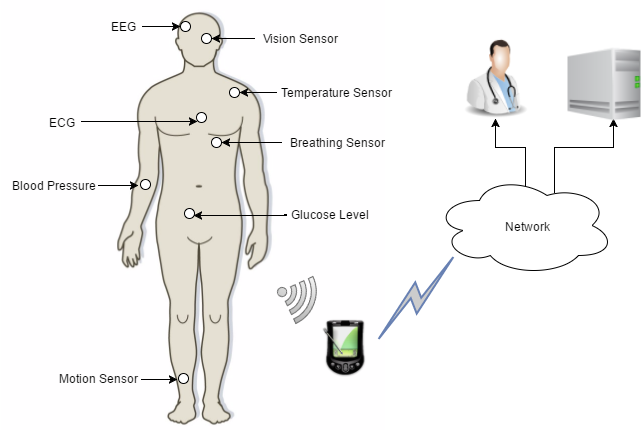}
\caption{WBAN General Architecture}
\label{fig:WBAN}
\end{figure}

The rapid advances in medical sensors enabled miniaturized computing devices called sensor nodes to be implanted in or placed around the human body. These sensors gather physiological and activity data from the human body and its environment, and send them wirelessly to a personal device like PDA (Personal Digital Assistant) or a smartphone that acts as a gateway to health care \cite{1, 17}. This network is called Wireless Body Area Network (WBAN) or Wireless Body Sensor Network (WBSN). A general architecture of WBAN is shown in Figure \ref{fig:WBAN}.

The main aim of WBAN is to enhance people’s life, thus many applications for WBAN can be found. In general, these applications can be categorized into medical and non-medical ones. Medical applications involve monitoring physiological attributes of the human body to detect any abnormal condition, allowing therefore appropriate personnel to take action before it is too late. Whereas examples of non-medical applications include entertainment applications, emotion detection, secure authentication, and non-medical emergencies through gathering data from the environment and warning people in case of danger like fire \cite{2}. WBAN are mostly used to monitor a single body but it can be extended to monitor a group of individuals known as collaborative WBAN systems \cite{3, 18, 19}.

In these Wireless Body Sensor Networks, sending data with adequate Quality of Service (QoS) like designing a high scalable system and sending data with a minimal delay is crucial. The delay and scalability of WBAN largely depend on the design and choice of the Medium Access Control (MAC) protocol.
Standard MAC protocols for WBANs include Static Time Division Multiple Access (Static TDMA), Dynamic Time Division Multiple Access (Dynamic TDMA), Frequency Division Multiple Access (FDMA), Carrier Sense Multiple Access with Collision Avoidance (CSMA/CA) and Direct Sequence Code Division Multiple Access (DS-CDMA). In the literature, many studies discuss the QoS characteristics of different MAC protocols \cite{5}, \cite{6}, \cite{7}, \cite{9}, \cite{10}, \cite{12}, \cite{13}, \cite{14}, \cite{16}. However, non of these researches compare the delay and the scalability characteristics of the five listed protocols simultaneously under the same experimental conditions in WBAN. In this study, each of the listed protocols is analyzed and compared with respect to two QoS metrics: delay and scalability.
The aim of this comparison is to show which technique offers the lowest delay and highest scalability in a high traffic environment. Such scenario can be induced in case of emergency for example, where physiological data collected from all sensors on human body should be sent simultaneously to properly assess the person's case and take action accordingly. This study can also be extended to cover collaborative WBAN systems where information from different bodies is sent simultaneously leading to high traffic.

This research work is organized as follows. A general review of basic MAC protocols for WBAN is presented in Section~\ref{sec:review}. A survey on the delay and the scalability performance of different MAC protocols is summarized in Section~\ref{sec:survey}. The experimental evaluation is presented in Section~\ref{sec:experiment}, while conclusion and future work are drawn in Section~\ref{sec:Conclusion and Future Work}. 

\section{General Review of MAC Protocols in WBAN}
\label{sec:review}
There are two main classes for MAC protocols: "contention-based" and "contention-free". In "contention-based" protocols, also known as "random access" protocols, nodes do not coordinate with each other to access the channel; so transmitted data may collide, forcing colliding nodes to backoff for a certain time before trying to access the channel again. Carrier Sense Multiple Access with Collision Avoidance (CSMA/CA) is an example of "contention-based" protocols. As for "contention-free" protocols, nodes follow a certain schedule to avoid collisions during transmission. Example of "contention-free" protocols include Static Time Division Multiple Access (Static TDMA), Dynamic Time Division Multiple Access (Dynamic TDMA), Frequency Division Multiple Access (FDMA), and Direct Sequence Code Division Multiple Access (DS-CDMA) \cite{4}.

\subsection{CSMA/CA}
\begin{figure}
\centering
\includegraphics[width=\columnwidth,height= 120mm,keepaspectratio]{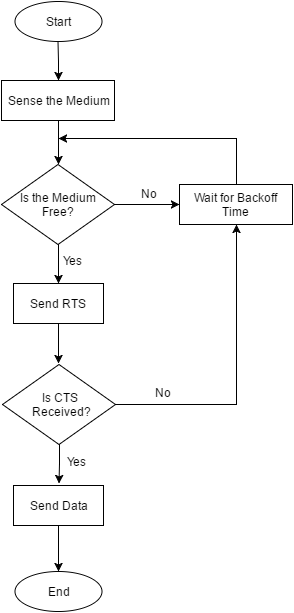}
\caption{CSMA/CA Algorithm}
\label{fig:CSMA}
\end{figure}

CSMA/CA is a random "contention-based" protocol. It is commonly known as the ON DEMAND access protocol since the sensor node access the transmission channel only when it has some information to send \cite{5}. Traditionally, CSMA nodes sense the medium prior to transmitting data. If the medium is free, they transmit the packet. However a “hidden problem” might occur in this case if another node is already sending data at the same time and will eventually result in collision \cite{6}. CSMA/CA is an enhancement over the traditional CSMA protocol in terms of collision avoidance capability. Improved CSMA/CA algorithm is shown in Figure \ref{fig:CSMA}. When a sensor node has data to send, it first senses the channel. If the channel is busy, the node waits for a random backoff time; once the channel is free, the node sends RTS (Request To Send) packet to the intended destination and waits to receive back a CTS (Clear to Send) packet. Receiving CTS indicates that it is safe to send information over the channel and therefore, data is transmitted to the destination. Otherwise, the sensor node goes to backoff time and waits till the channel is free again \cite{6}.

\subsection{Static TDMA}
\begin{figure}
\centering
\includegraphics[width=\columnwidth,height=\textheight,keepaspectratio]{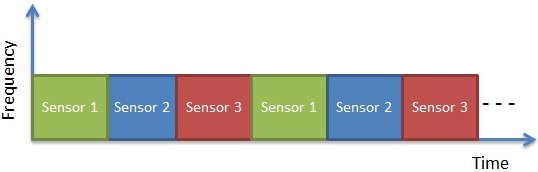}
\caption{Static TDMA Access Scheme}
\label{fig:Static TDMA}
\end{figure}

Static Time Division Multiple Access (Static TDMA) is a scheduled "contention-free" protocol in which the time frame is divided into dedicated time slots. Every slot is assigned to a sensor node and each node sends data in succession one after another during its corresponding slot \cite{7}. The Static TDMA access scheme is presented in Figure \ref{fig:Static TDMA} \cite{8}.

\subsection{Dynamic TDMA}

\begin{figure}
\centering
\includegraphics[scale=0.6]{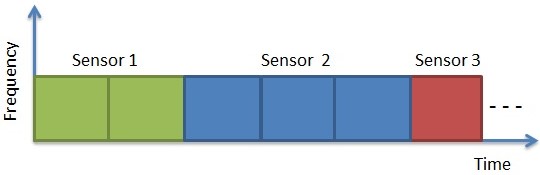}
\caption{Dynamic TDMA Access Scheme}
\label{fig:D-TDMA}
\end{figure}

In Dynamic Time Division Multiple Access (Dynamic TDMA), also known as Reservation-Based TDMA, a variable number of time slots is dynamically reserved to different nodes using a scheduling algorithm based on the traffic demand of each data stream. Slots are therefore reserved to the nodes encountering high traffic (buffered packets) and are released from other nodes after completing the data transmission and reception \cite{9}. The Dynamic TDMA access scheme is shown in Figure \ref{fig:D-TDMA}.

\subsection{FDMA}
\begin{figure}
\centering
\includegraphics[scale=0.7]{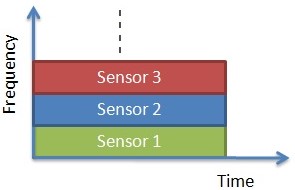}
\caption{FDMA Access Scheme}
\label{fig:FDMA}
\end{figure}

FDMA is another "contention-free" protocol in which nodes are assigned different frequency bands to transmit their data through the medium. Each frequency band is separated from its adjacent bands by a guard band to avoid interference. Therefore, nodes can transmit their data without any need for further process \cite{6}. The FDMA access scheme is shown in Figure \ref{fig:FDMA} \cite{8}.

\subsection{DS-CDMA}
\begin{figure}
\centering
\includegraphics[scale=0.7]{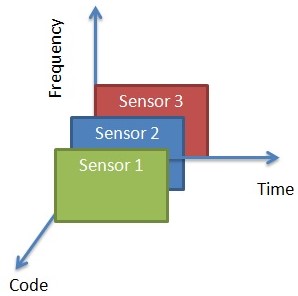}
\caption{DS-CDMA Access Scheme}
\label{fig:CDMA}
\end{figure}

In Direct Sequence Code Division Multiple Access (DS-CDMA), every node is assigned a unique code \cite{10}. All nodes send their data over the same frequency, but they are still distinguished from one another by the different codes assigned to them. The user’s generated code is multiplied with the user’s original signal to form his encoded signal, \textit{i.e.}, encoded signal = (original signal) x (code). Hence the nomination Direct Sequence-CDMA or DS-CDMA. The DS-CDMA access scheme is shown in Figure \ref{fig:CDMA} \cite{8}.

\section{Delay and Scalability Performance Survey}
\label{sec:survey}

Reliability in WBAN is very important, and end-to-end delay is a major key performance metric in critical healthcare applications \cite{11}. Studying the delay performance of different MAC protocols is specifically important in emergency cases requiring simultaneous transmission of critical data from all sensors in or on the body to take fast action accordingly. Scalability or system capacity is another important QoS metric in WBAN since sensor nodes might be added to the system at anytime based on the medical need or on the type of physiological data that should be gathered. Scalability reflects the performance of the MAC protocol when more sensor nodes are added to the system and shows how flexible it is to add them. As mentioned earlier, many articles discuss the QoS characteristics of MAC protocols in WBAN, but non of these researches compare the five protocols concurrently in the same environment. In the following, we combine the findings from different studies analyzing the delay and scalability performance for Static TDMA, Dynamic TDMA, FDMA, CSMA/CA and DS-CDMA.
\subsection{Delay Analysis}

In \cite{12}, the authors present a comparison between FDMA and Static TDMA techniques. They demonstrate that in general, Static TDMA induces less delay than FDMA since the transmission of a TDMA packet takes only one slot, whereas the transmission in FDMA lasts for a whole frame. Also, authors prove that the difference in delay between the two protocols is variable since in Static TDMA, a packet has to wait for its appropriate slot even when the queue is empty, whereas the packet is instantly transmitted without further delay in FDMA. Nevertheless, this same literature proves that when the load increases, the ratio of the delays of both schemes become close to one, so Static TDMA and FDMA will have similar performances. Authors of \cite{6} state that in Static TDMA, the generated packets experience three types of delays before reaching the receiver: transmission delay, queuing delay, and propagation delay. In this literature, a comparison of delay as a function of throughput in WBAN is assessed for various protocols including Static TDMA, FDMA, and CSMA/CA. The comparison shows that in low traffic, both Static TDMA and FDMA offer low delay, and Static TDMA outperforms FDMA. However the delay significantly increases in both schemes when the load increases. As for CSMA/CA, the delay is the highest among the other protocols even when the traffic is low since CSMA/CA continuously senses the medium and waits for it to become free before transmitting the packets. Literature \cite{7} examines the delay characteristics of different protocols used in Wireless Sensor Networks under varying traffic loads. It shows that the average message delay of Static TDMA increases with increasing traffic load due to queuing that is originated from the limited bandwidth available since every node is transmitting one message per frame. This literature also explains that the delay induced by CSMA/CA is due to two factors: the "contention-based" nature of CSMA/CA when the node renounces from sending its data after finding the channel busy, and the retransmission of the messages due to collision. Authors of \cite{13} analyze the performance of DS-CDMA and Static TDMA when bursty voice traffic is sent, and prove that even though Static TDMA outperforms DS-CDMA in low traffic, the delay of DS-CDMA when bursty data is applied is much lower than that of Static TDMA. Literature \cite{14} compares DS-CDMA based protocols to other protocols in Wireless Sensor Networks and states that the delay of DS-CDMA is induced by assigning different codes for every node. Authors of \cite{10} introduce a Dynamic TDMA scheme for WBAN, and show that the delay in Dynamic TDMA is lowest delay realizable by TDMA mechanism.

Table \ref{table:1} summarizes the protocols studied in every reference in terms of delay. It shows that none of the references compared the delay performance of the five protocols simultaneously. Table \ref{table:2} presents the delay analysis results based on the literature findings.


\begin{table*}[!h]
\renewcommand{\arraystretch}{1.3}
\centering
\begin{tabular}{||c|| c| c| c| c| c||} 
 
\hline
 Reference & Static-TDMA & Dynamic-TDMA & FDMA & CSMA/CA & DS-CDMA \\ [0.5ex]
 \hline\hline
 \cite{6} & yes & no & yes & yes & no \\  \hline
 \cite{7} & yes & no & no & yes & no \\  \hline
 \cite{10} & yes & yes & no & no & no \\   \hline
 \cite{12}& yes & no & yes & no & no \\  \hline
 \cite{13} & yes & no & no & yes & yes\\ \hline
 \cite{14} & no & no & no & no & yes\\

 \hline
\end{tabular}

\caption{ MAC PROTOCOLS STUDIED IN LITERATURE - DELAY ANALYSIS}
\label{table:1}
\end{table*}

\begin{table*}[!h]

\renewcommand{\arraystretch}{1.3}
\centering
\resizebox{130mm}{!}{
\begin{tabular}{||c|| c| c| c| c| c||} 
 \hline
 Performance Metric & Static-TDMA & Dynamic-TDMA & FDMA & CSMA/CA & DS-CDMA \\ [0.5ex]
 \hline
 Delay & High & High & High & High & Low \\  
 \hline
\end{tabular}
}
\caption{DELAY ANALYSIS RESULTS}
\label{table:2}
\end{table*}

\subsection{Scalability Analysis}
Many articles show that scalability is poor in TDMA based system. For instance, authors of \cite{5} state that scalability is a main disadvantage in TDMA based systems since adding a sensor node requires performing modifications in the central controller. For example, if a medical staff decided to add sensor nodes to monitor additional physiological data, he has to change the transmission time frame from the central controller which is not practical. This literature suggests that TDMA is suitable for small WBAN systems with limited number of nodes. Also, authors of \cite{10}, \cite{14} and \cite{9} state that TDMA has limited scalability and ability to adapt to changes such as adding sensor nodes. Literature \cite{12} shows that in both FDMA and Static TDMA, the delay increases with the number of users. So the performance of both protocols degrades when increasing the number of nodes which reflects a poor scalability. Authors of \cite{6} show that CSMA/CA has good scalability in WBAN, as it maintains a constant delay when increasing the offered load. Also, literature \cite{5} and \cite{10} state that scalability is one of the important advantages of CSMA/CA as it can easily accommodate different traffic sources with different rates. Article \cite{16} states that capacity is a main advantage of DS-CDMA as it can handle more nodes than the other technologies.

Table \ref{table:3} summarizes the protocols studied in every reference in terms of scalability. It shows that none of the references compared the scalability performance of the five protocols simultaneously. Table \ref{table:4} presents the scalability analysis results based on the literature findings.

\begin{table*}[!h]
\renewcommand{\arraystretch}{1.3}
\centering
\begin{tabular}{||c|| c| c| c| c| c||} 
 \hline
 Reference & Static-TDMA & Dynamic-TDMA & FDMA & CSMA/CA & DS-CDMA \\ [0.5ex] 
 \hline\hline
 \cite{5} & yes & yes & no & yes & no \\  \hline
 \cite{6} & yes & no & yes & yes & no \\  \hline
 \cite{9} & yes & no & no & no & no \\  \hline
 \cite{10} & yes & no & no & yes & no\\  \hline
 \cite{12} & yes & no & yes & no & no \\  \hline 
 \cite{14} & yes & no & no & no & no \\  \hline
 \cite{16} & no & no & no & no & yes \\ 
 \hline
\end{tabular}
\caption{MAC PROTOCOLS STUDIES IN LETERATURE - SCALABILITY ANALYSIS}
\label{table:3}
\end{table*}

\begin{table*}[!h]
\renewcommand{\arraystretch}{1.3}
\centering
\resizebox{131mm}{!}{
\begin{tabular}{||c|| c| c| c| c| c||} 
 \hline
 Performance Metric & Static-TDMA & Dynamic-TDMA & FDMA & CSMA/CA & DS-CDMA \\ [0.5ex] 
 \hline\hline
 Scalability & Poor & Poor & Poor & Good & Good \\ 
 \hline
\end{tabular}
}
\caption{SCALABILITY ANALYSIS RESULTS }
\label{table:4}
\end{table*}

\section{Experimental Evaluation}
\label{sec:experiment}

\subsection{Motivations}

To fairly assess the performance of Static TDMA, Dynamic TDMA, FDMA, DS-CDMA and CSMA/CA protocols, the five techniques should be tested under the same experimental conditions, which has not been done previously as shown in Table \ref{table:1} \& Table \ref{table:3}. Therefore in this section, delay and scalability will be simulated simultaneously for all the listed protocols in the same high traffic environment. Simulation results will be compared to those presented in Table \ref{table:2} \& Table \ref{table:4} in order to draw appropriate conclusion about the best protocol to be used in a high traffic environment.



\subsection{Simulation Environment and Parameters}
\begin{figure}[!h]
\centering
\includegraphics[height=75mm,width=75mm]{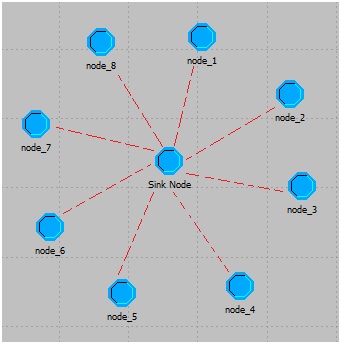}

\caption{OPNET Star Topology}
\label{fig:star}
\end{figure}

Testing the performance of the five listed MAC protocols is done using OPNET simulator \cite{20}. In the simulation, eight nodes are placed in a star topology architecture around the sink node as shown in Figure \ref{fig:star}.

The OPNET default parameters for the physical layer and wireless link were used. The transmission rate is set to 250kbps, and the packet arrival time between packets is set to 5ms in order to induce high traffic.  The simulation parameters are summarized in Table \ref{table:5}.

\begin{table*}[!h]
\renewcommand{\arraystretch}{1.3}
\centering
\resizebox{55mm}{!}{
\begin{tabular}{||l|l ||} 
 \hline
 Parameter & Value\\ [0.5ex]
 \hline\hline
 
 Number of Nodes & 8 \\ \hline
 Packet Size & 54 bytes \\  \hline
 Packet Arrival Time & 5ms \\ \hline
 Bit Rate & 250kbps\\ \hline
 Distance Between Nodes & 2m\\ [1ex] 
 \hline
\end{tabular}
}
\caption{SIMULATION PARAMETERS }
\label{table:5}
\end{table*}

\subsection{Obtained Results and Discussion}
\begin{figure}[!h]
\centering
\includegraphics[width=\columnwidth,height=56mm]{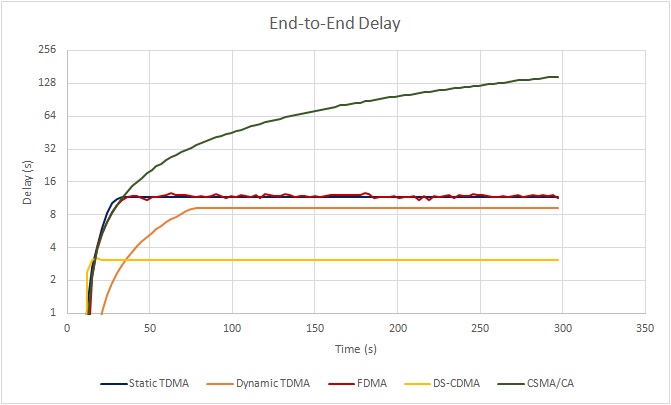}
\caption{Delay Performance of MAC Protocols}
\label{fig:delay}
\end{figure}

The delay simulation results are presented in Figure \ref{fig:delay}. They show that DS-CDMA outperforms the other "contention-free" protocols since it induces the lowest delay. This delay is generated from assigning different codes to nodes. Static TDMA and FDMA are very close in performance in a high traffic environment as they both induce high delays due to queuing. The delay of Dynamic TDMA is high, but less than that of Static TDMA since it uses the slots more efficiently, and therefore minimizes queuing. As for CSMA/CA, the "contention-based" protocol, the delay is unmanageable since in high traffic, the probability of collision in CSMA/CA increases, so the number of retransmissions accumulate. Also, the medium is very busy, therefore the nodes have to wait for a long backoff time. The obtained results agree with the delay analysis results obtained in Table \ref{table:2}.

\begin{figure}[!h]
\centering
\includegraphics[width=\columnwidth,height=\textheight,keepaspectratio]{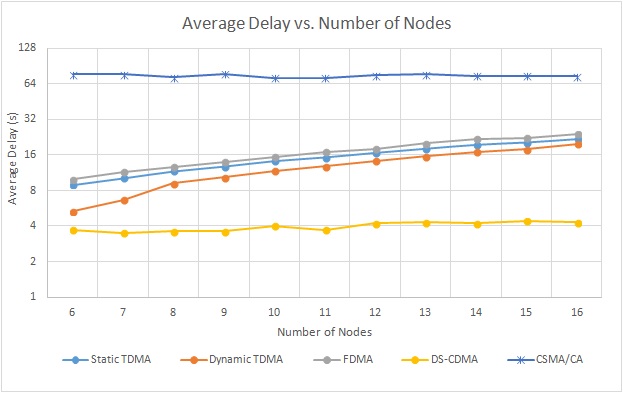}
\caption{Scalability Performance of MAC Protocols}
\label{fig:scal}
\end{figure}

In order to test the scalability of different protocols, simulations are repeated with different number of nodes, and the average end-to-end delay induced by each protocol was evaluated accordingly. Figure  \ref{fig:scal} shows the scalability performance of each protocol with respect to different number of nodes.

Results show that the delay in Static TDMA, Dynamic TDMA and FDMA increases with the number of nodes. For instance, the average end-to-end delay of Static TDMA increased from 8.93 seconds to 21.6 seconds when the number of nodes is doubled from eight to sixteen. Similarly, the average delay of Dynamic TDMA increased from 5.35 seconds to 19.77 seconds, and that of FDMA increased from 9.91 seconds to 23.85 seconds. This implies a degradation in the performance of these three protocols when more nodes are added to the system, and reflects a poor scalability. On the other hand, the delay in DS-CDMA and CSMA/CA protocols remains almost constant when increasing the number of nodes, therefore these two protocols offer good scalability. The results obtained in the simulation agree with the scalability analysis results presented in Table \ref{table:4}.

Both the delay and scalability studies showed that DS-CDMA MAC protocol outperforms the other protocols when high traffic is generated in WBAN since it offers the lowest delay and the highest scalability.




\section{Conclusion and Future Work}
\label{sec:Conclusion and Future Work}


In this paper, the delay and scalability performance of Static TDMA, Dynamic TDMA, FDMA, DS-CDMA and CSMA/CA MAC protocols were analyzed and simulated in a high traffic WBAN environment. Simulation results showed that DS-CDMA protocol performs better than the other techniques in terms of delay and scalability, which agrees with the survey analysis. Therefore DS-CDMA should be considered to be used in high traffic WBAN systems. Future work includes analyzing and comparing the performance of these five protocols in terms of other QoS metrics like probability of collision, hardware complexity and energy consumption for better assessment. In addition, the study presented in this paper is based on a single body WBAN, so it can be extended to analyse the behavior of the different protocols in collaborative WBAN systems.


\printbibliography

\end{document}